\documentclass[preprint2]{aastex}

\usepackage{natbib}
\usepackage{graphicx}
\usepackage{epstopdf}

\shorttitle{New Burst Class Discriminator}
\shortauthors{Goldstein, Preece \& Briggs}

\begin{document}

\title{A New Discriminator for Gamma Ray Burst Classification: \\
    The $E_{\rm peak}$ -- Fluence Energy Ratio}

\author{Adam Goldstein, Robert~D.~Preece and Michael~S.~Briggs}
\affil{Physics Department, The University of Alabama in Huntsville,
    Huntsville, AL 35809}

\begin{abstract}
Using the derived gamma-ray burst $E_{\rm peak}$ and fluences from the complete BATSE 5B Spectral 
Catalog, we study the ensemble characteristics of the $E_{\rm peak}$--fluence relation for 
GRBs.  This relation appears to be a physically meaningful and insightful fundamental discriminator 
between long and short bursts.  We discuss the results of the lower limit test of the 
$E_{\rm peak}$--$E_{\rm iso}$ relations in the $E_{\rm peak}$--fluence plane for BATSE 
bursts with no observed redshift.  Our results confirm the presence of two GRB classes as well 
as heavily suggesting two different GRB progenitor types.
\end{abstract}

\keywords{gamma rays: bursts --- methods: data analysis}

\section{Introduction}

Classification of Gamma-Ray Bursts (GRBs) is certainly a difficult task. 
Bursts are divided into long and short classes, based upon the bimodal duration 
histogram of bursts \citep{Kouveliotou} observed by the Burst And Transient Source 
Experiment (BATSE), which was on board the Compton Gamma Ray Observatory. One parameter, 
the split time of 2 seconds on the $t_{90}$ durations plot \citep{Koshut}, was sufficient to classify 
bursts.  Bursts from the two classes have another discriminator, spectral hardness as determined 
by the ratio of two broad energy channels.  This hardness ratio, when used in 
conjunction with the duration, provides a means for classification, as shown by \citet{Kouveliotou}.  
In addition, another classification scheme uses the scatter plot of the fluence and duration fitted  
with two 2D Gaussians \citep{Balazs}. Some have indicated there are more than two clusters 
\citep{Mukherjee, Horvath}, while \citet{Hakkila} maintain that the reported third cluster is simply the 
result of a selection bias.  Furthermore, significant overlap is present in the duration comparison of 
short and long bursts, complicating a clear distinction between two classes as has been discussed 
by the authors cited above.  In any case, there are difficulties with all aforementioned classification 
measures in that the $t_{90}$ duration is somewhat subjective in the necessary selections of 
background regions, while a hardness ratio based upon counts is strongly detector dependent.

The observation that some classically short bursts are extended in duration, when observed in
an energy band (BATSE 20-50 keV) different from that of the natural BATSE 50-300 keV band, 
was reported by \citet{Lazzati}.  The dedicated GRB mission, Swift \citep{Barthelmy}, has 
introduced additional issues, while reinforcing the extended emission of short GRBs at lower 
energies.  \citet{Norris} have introduced the time lag between broad energy channels (`spectral 
lag') as a classifier: `short' GRBs have approximately zero lag, while the `long' events have 
a lag that is significantly different from zero. Indeed, having a near-zero lag is the
basis for claiming that some bursts observed by the Swift BAT that are significantly longer than 2
seconds belong in the `short' class \citep{longShorts, Zhang}.

Following on the analysis of \citet{NakarPiran} on the Amati relation \citep{Amati02}, 
\citet{BandPreece} investigated the implications of combinations of several observable 
GRB parameters, derived from an extensive data set of GRBs observed 
by BATSE. The BATSE data set used was a partial spectral catalog of the peak flux 
and fluence spectral parameters \citep{Mallozzi}, complete up to the end of the BATSE 4B 
Catalog \citep{Paciesas}. The BATSE 5B Spectral Catalog has now been completed 
\citep{5BCatalog}, which includes all BATSE bursts with sufficient counts 
in the spectral data that they could be analyzed. Based on this comprehensive data set, we 
present a new GRB classification measure that is as diagnostic as the $t_{90}$ duration for 
classification, but does not rely on the subjective choices required for the durations calculation 
\citep{Koshut}.

\section{Motivation}
BATSE data has been used to study the $E_{\rm peak}$ distributions of GRBs 
\citep{Kaneko}, and the time-integrated $E_{\rm peak}$ distribution for all bursts shows no 
evidence of discrimination between short and long bursts.  Fluence hardness distributions 
\citep{Kouveliotou} show some evidence of bimodality \citep{Balazs}, but there is only 
moderate significance with much overlap \citep{Nakar}.  Using the BATSE 5B Spectral Catalog, we 
support previous observations on the distributions of $E_{\rm peak}$ and fluence.  In addition, we 
split the distributions into long and short duration GRB distributions, following the $t_{90}$ 
classification of two seconds.  The $E_{\rm peak}$ distribution for short bursts is completely 
overlapped by that of long bursts.  Although the $E_{\rm peak}$ values for long bursts are centered 
around 150 keV, short burst $E_{\rm peak}$ values are shifted to higher energies around 300 keV.
This is consistent with previous findings of \citet{Paciesas2} and \citet{Ghirlanda2}.
Approximately 65\% of the fluence distribution for short bursts overlaps that of the fluence 
distribution for long bursts, with the position of the peak of the short GRB fluence distribution being an order of 
magnitude less than that for long GRBs.  

\citet{LloydPetrosian} have shown there to be a significant correlation between 
$E_{\rm peak}$ and the total fluence in gamma-rays, and it is desirable to investigate this 
correlation for both long and short GRBs.  For this reason, we investigate the $E_{\rm peak}$--
fluence distribution for all BATSE bursts with good spectral fits and devise a discriminator between 
long and short bursts based on the difference in correlation between $E_{\rm peak}$ and fluence 
for long and short bursts.  A choice formulation for a discriminator is based on the hardness 
of a burst.  \citet{Kouveliotou} used a hardness ratio based on the ratio of calculated fluence in 
different energy bands to compare to duration estimates.  Instead, we propose to use a hardness 
measure represented by $E_{\rm peak}$/fluence.  This so-called energy ratio is in units of area and 
should prove to be a good discriminator between long and short bursts if there is a 
strong correlation between $E_{\rm peak}$ and fluence.

\section{Observations}
From the 5B spectral catalog (Goldstein, Preece \& Mallozzi, in prep.), 
we extract bursts with a good model fit as determined by a 3-sigma 
confidence limit, with the time-integrated $E_{\rm peak}$ and fluence errors for each burst 
required to be no more than 40\% of their respective fitted values.  A total of 1121 long bursts 
and 168 short bursts, classified according to the classical $t_{90}$ cut of two seconds 
\citep{Kouveliotou}, satisfied these criteria.  We then calculated the $E_{\rm peak}$/fluence energy 
ratio for each of these bursts and plot a histogram of these values, as in Fig. 1.  By using a standard
nonlinear least-squares fitting algorithm, we fit a single lognormal function to the distribution with the
resulting chi-square goodness-of-fit statistic 111 for 32 degrees of freedom.  We then fit two lognormal
functions to the distribution with the resulting chi-square statistic of 32 for 29 degrees of freedom.  
Since the two models are nested, we use Pearson's chi-square test to show that the large change 
in chi-square per degree of freedom results in a chance probability of $5 \times 10^{-17}$  and that the 
two lognormals are statistically preferred with a high degree of significance .  

From this bimodal distribution an obvious distinction between long and short bursts emerges.  In Fig. 2
we plot two histograms corresponding to long and short bursts as identified by their respective $t_{90}$
estimations to show that the bimodal distribution of the energy ratio is correlated to that of the duration
distribution.  A K-S test comparing the long and short burst distributions in Fig.~2 to the best fit lognormal
functions in Fig.~1, however, finds the correlations to be statistically marginal with 2\% and 0.6\% respective
probabilities that each distribution in Fig.~2 is drawn from their respective best fit lognormal functions in Fig.~1.  The 
energy ratio distribution for short bursts is narrower compared to that of long bursts and
is shifted to higher energies, resulting from the fact that short bursts are generally harder than long bursts.  
It appears the energy ratio values are a good discriminator between the classical definition of long and short 
bursts by merging two known discriminators into one quantity, and their relative overlap can be well 
estimated.  Only 4\% of long GRBs overlap the 1-sigma core of the short burst distribution, and 2\% of short bursts
overlap the the 1-sigma core of the long burst distribution.  Similarly, the overlap for the 2-sigma cores is
11\% and 23\%, respectively.  Comparatively, for our sample, the classical $t_{90}$ overlap of long 
bursts onto the 1-sigma (2-sigma) core of short bursts is 4\% (23\%), and and there is no overlap of 
short bursts onto either the 1-sigma or 2-sigma core of long bursts.  The central value for the long 
burst energy ratio distribution is $\sim$0.06 while the central value for the short burst distribution is $\sim$1.5.

\citet{BandPreece} showed that the Amati relation \citep{Amati02},
\begin{equation}
E^{\rm rest}_{\rm peak} = C_{\rm A} \ \Biggl( \frac{E_{\rm iso}} {10^{52} \ \rm{erg}} \Biggr)^{\eta_{\rm A}}
\end{equation}
and the Ghirlanda relation \citep{Ghirlanda}, 
\begin{equation}
E^{\rm rest}_{\rm peak} = C_{\rm G} \ \Biggl( \frac{E_{\rm iso} \ f_{\rm B}} {10^{51} \ \rm{erg}} \Biggr)^{\eta_{\rm G}},
\end{equation}
could be converted into a similar energy ratio 
\begin{equation}
\frac{E^{1/\eta_i}_{\rm peak, obs}}{S_\gamma} \propto F(z),
\end{equation}
where the $C_{\rm i}$ in the previous equations are the respective normalization coefficients and 
$f_{\rm B}$ is the beaming fraction relevant for the Ghirlanda relation.
Here, $S_\gamma$ is the fluence in gamma-rays, and $\eta_i$ are the best fit power law indices for 
the respective models.  These energy ratios can be represented as functions of redshift, $F(z)$, and the 
upper limit of the ratios could be determined for any redshift.  The energy ratio upper limit of the 
Amati relation, as well as the energy ratio upper limit of the Ghirlanda relation, can be projected 
into the $E_{\rm peak}$ -- fluence plane where they become lower limits.  Using the bursts  
described above, the $E_{\rm peak}$ values and fluences can be plotted in this plane.  Fig. 3 
shows the distribution of long bursts in the upper plot and the distribution of short bursts in the 
bottom plot.  The lines denote the lower limits of the Amati and Ghirlanda relations in this plane.  
Note that the beaming fraction for the Ghirlanda relation, $f_{\rm B}$ = 1.0, which is related to the jet 
opening angle, $\theta$, by $f_{\rm B} = 1- \cos{\theta}$.  This represents the energy radiated by the 
burst equalling $E_{\rm iso}$.  It can easily be seen there are two separate distributions in the 
$E_{\rm peak}$ -- $E_{\rm iso}$ plane.  Long bursts appear to be clumped between the Amati and 
Ghirlanda limits, while the short bursts appear to distribute along the Ghirlanda lower limit. 

\section{Conclusions}
The energy ratio shows a clear distinction between two different types of GRBs.  The fluence encodes 
the duration of the burst without deriving a subjective $t_{90}$ estimate, and $E_{\rm peak}$/fluence 
physically represents a ratio of the energy at which most of the gamma-rays are emitted to the total 
energy emitted in gamma-rays.  This quantity effectively serves as a spectral hardness ratio and 
shows an increased hardness for short bursts compared to long bursts, consistent with \citet{Kouveliotou}.  
The distribution of short bursts is narrower, its 1-sigma width covering less than one order of magnitude in 
energy, while the long bursts 1-sigma width covers slightly more than an order of magnitude.  
This bimodal distribution heavily supports the original distinction between long and short bursts 
\citep{Kouveliotou} and suggests further investigation is desirable.  The correlation between 
$E_{\rm peak}$ and the total fluence in gamma-rays is of particular interest, as the energy ratio 
removes the cosmological dependence for the energies involved since its value is merely proportional 
to the square of the luminosity distance.  Because of this reason, the energy ratio could be considered 
physically superior to the duration classifier for GRBs.  The low degree of correlation between the $t_{90}$
distribution and the energy ratio distribution may be attributed to the fact that the former is a an observed
quantity in the observer's frame, while the latter is a quantity that contains spectral information from the 
rest frame of the GRB.  In addition, when comparing the difference in overlap between the energy ratio distributions 
and the $t_{90}$ distributions, the overlap for the $t_{90}$ distributions is marginally less pronounced, but
gives little insight to the physical processes in the rest frame of the GRB.   Note that there may be 
a truncation effect with the energy ratio associated with short bursts, due to the inability of BATSE to trigger 
on very low fluence events and bursts with $E_{\rm peak}$ outside the BATSE energy range.  It is expected 
that Fermi/GBM will assist in discovering GRBs with $E_{\rm peak}$ values greater than 1 MeV, alleviating 
possible truncation effects due to the energy cutoff at the high end of the detector energy band.  

While a majority of BATSE bursts ($\sim$87 percent) fail the lower limit test for the Amati relation
in the $E_{\rm peak}$ -- fluence plane, very few BATSE bursts violate the lower limit for the 
Ghirlanda relation with $f_{\rm B}$ = 1.0.  Especially intriguing is the fact that short GRBs fall close to the
lower limit where the energy radiated is isotropic.  Decreasing the beaming fraction shifts the lower
limit to higher fluences, causing an increasing number of bursts to violate the Ghirlanda relation.
In addition, only a few jet breaks have been discovered for short GRBs \citep{Soderberg}, and 
\citet{Watson} caution those discoveries may be misleading due to the flaring activities of short 
GRB decay.  Thus, the Ghirlanda relation, as well as the lack 
of reliable observed jet breaks for short bursts, suggests that short bursts release energy 
isotropically (or near isotropically), while longer bursts tend to have a much smaller beaming 
fraction and consequently have small opening angles.   Clearly the Ghirlanda relation, if accurate, 
requires short bursts to release energy isotropically, as opposed to beamed radiation release in 
long bursts \citep{Frail, Nakar}.  In addition, if short GRBs are isotropic emitters, then virtually all 
short GRBs should be detectable within a given volume of space.  Long GRBs, in general, have a 
small measured opening angle, and therefore only a small fraction are detectable \citep{Paradijs}.  
Since short GRBs are detected far less frequently than long GRBs \citep{Paciesas} this is indicative 
of a relatively rare and independent cause for short GRBs.


\clearpage

\begin{figure}
\includegraphics[scale=.5, angle=270]{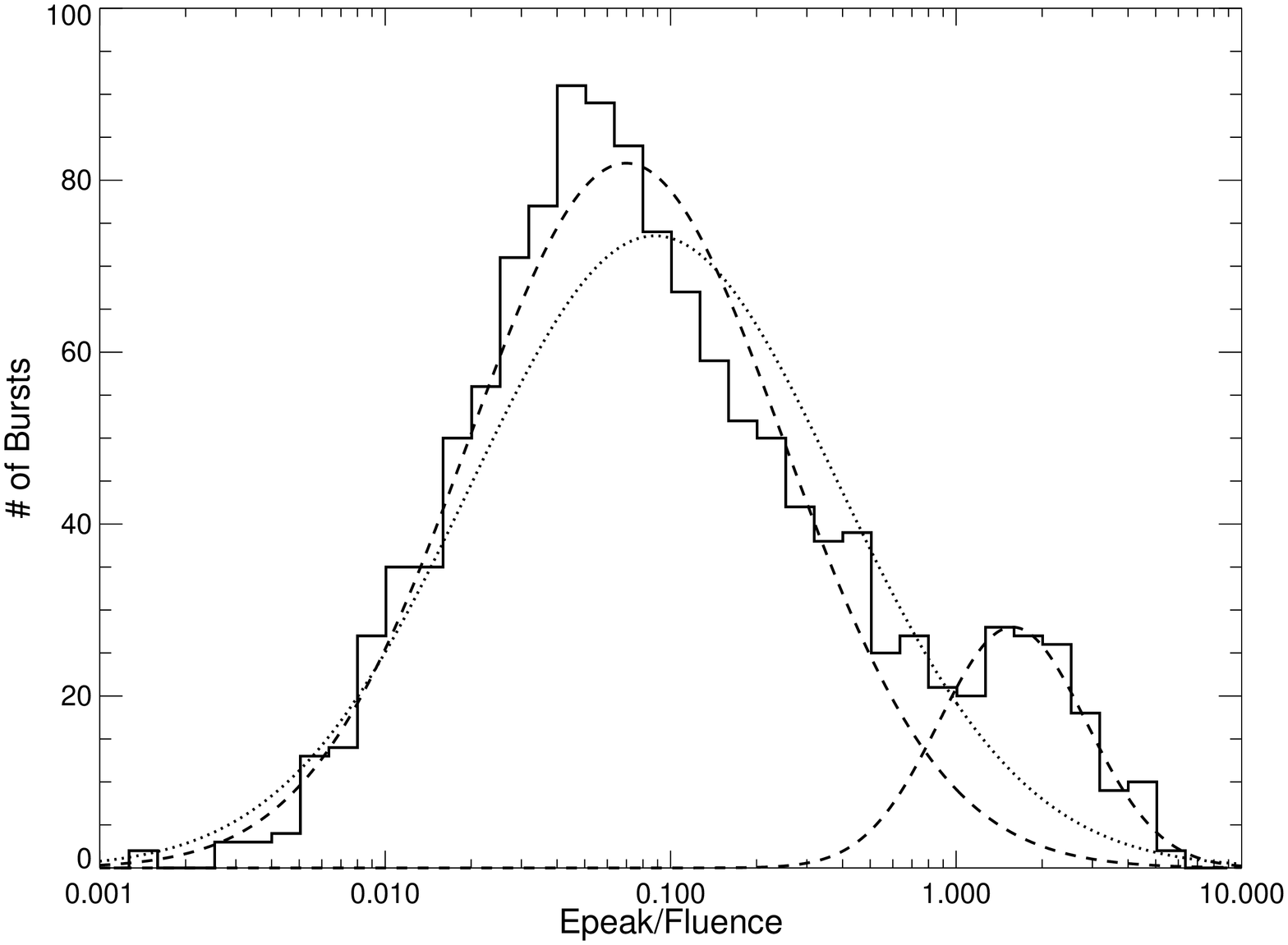}
\caption{Histogram of the energy ratio distribution in the 20-2000 keV range 
for 1289 GRBs.  The dotted line shows the best fit lognormal to the distribution, and the dashed lines show
the fit of two lognormal functions.  The chi-square goodness-of-fit statistic for one lognormal is 111 for 32 
degrees of freedom, while for two lognormals is 32 for 29 degrees of freedom.  Therefore, two lognormal 
distributions are statistically perferred, resulting in a bimodal distribution for the energy ratio. \label{fig1}}
\end{figure}

\clearpage

\begin{figure}
\includegraphics[scale=.5, angle=270]{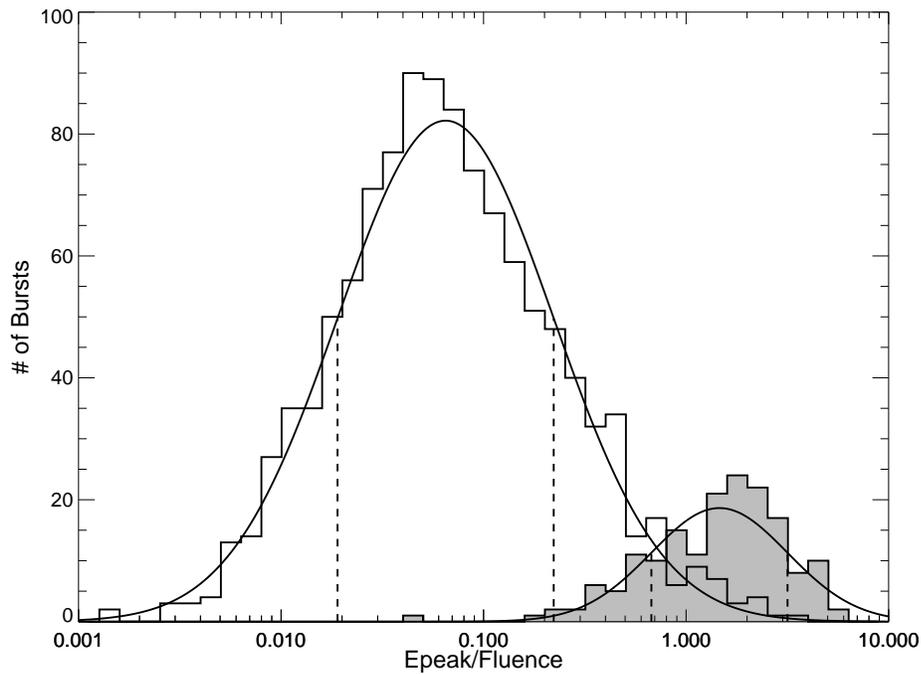}
\caption{Histograms of the energy ratio distributions in the 20-2000 keV range 
for 1121 long bursts (white) and 168 short bursts (gray).  There are clearly two distinct distributions, 
with long bursts centered around 0.6 and short bursts centered around 1.5.  The solid curves are the best fit 
lognormal functions, and the dashed lines are the 1-sigma standard deviation of the distributions.  \label{fig2}}
\end{figure}

\clearpage

\begin{figure}
\includegraphics[scale=.5]{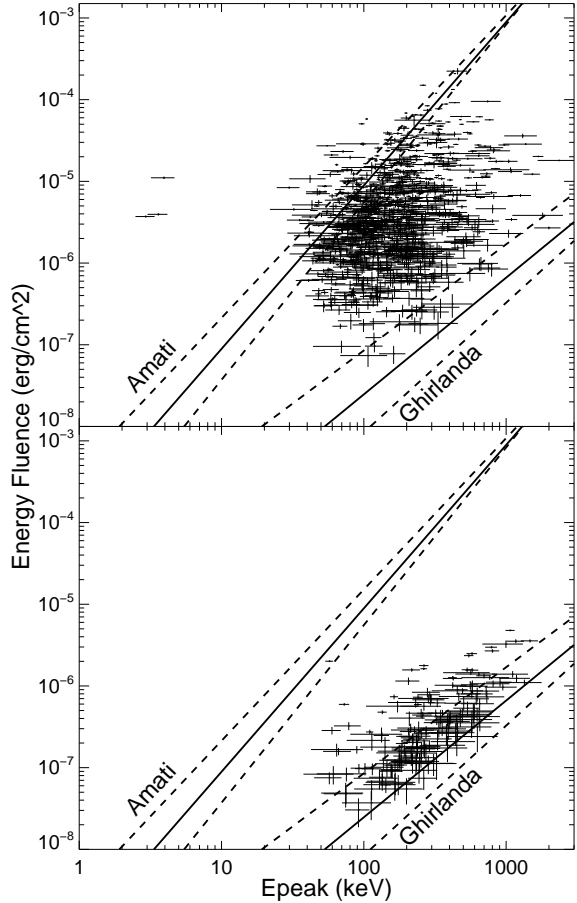}
\caption{Scatter plots of 1121 long bursts (top) and 168 short bursts (bottom) in the 
$E_{\rm peak}$--fluence plane.  Also plotted are the lower limits of the Amati relation and 
Ghirlanda relation ($\rm{f}_B$=1.0).  The dashed lines represent the 1-sigma errors about the 
lower limits.  There is apparently a clear lower limit violation by most BATSE bursts for the Amati 
relation.  Short bursts appear to cluster around the lower limit of the 
Ghirlanda relation. \label{fig3} }
\end{figure}

\end{document}